\documentclass[epj]{svjour}
\usepackage{latexsym}
\usepackage{graphics}
\def\dfrac#1#2{{\displaystyle {#1 \over #2}}}

\def\slash#1{\setbox0=\hbox{$#1$}\dimen0=\wd0                    
      \setbox1=\hbox{/} \dimen1=\wd1 \ifdim\dimen0>\dimen1
      \rlap{\hbox to \dimen0{\hfil/\hfil}} #1                        \else                                       
      \rlap{\hbox to \dimen1{\hfil$#1$\hfil}}  
      /   \fi}                                         

\newcommand{\be}{\begin{equation}}
\newcommand{\ee}{\end{equation}}
\newcommand{\bea}{\begin{eqnarray}}
\newcommand{\eea}{\end{eqnarray}}
\newcommand{\nn}{\nonumber}

\newcommand{\Heff}{{\cal H}_{ eff}}

\newcommand{\DB}{\Delta B}

\begin{document}
\title{Beauty Hadron Lifetimes and B-Meson CP-Violation Parameters from Lattice QCD}
\author{Cecilia Tarantino\inst{}
\thanks{\emph{Present address:} tarantino@fis.uniroma3.it}%
}                     % Do not remove
\offprints{}          % Insert a name or remove this line
\institute{Dip. di Fisica, Univ. di Roma Tre and INFN, Sezione di Roma III, Via della Vasca Navale 84, I-00146 Rome, Italy}
\date{Received: date / Revised version: date}
\abstract{
The present status of the theoretical estimates of beauty hadron lifetime ratios and of width differences and CP-violation parameters in $B_d$ and $B_s$ systems is reviewed.
In the last two years accurate lattice calculations and next-to-leading order perturbative computations have improved these theoretical predictions, leading to the following updated results: $\tau(B^+)/\tau(B_d)=
1.06 \pm 0.02, \,  
\tau(B_s)/\tau(B_d)=
1.00 \pm 0.01, \, 
\tau(\Lambda_b)/\tau(B_d)=
0.88 \pm 0.05, \,\Delta \Gamma_{d}/\Gamma_{d} = (2.42 \pm 0.59)10^{-3}, \,
\Delta \Gamma_{s}/\Gamma_{s} = (7.4 \pm 2.4)10^{-2}, \,\vert(q/p)_d\vert-1=(2.96 \pm 0.67) 10^{-4}$ and $\vert(q/p)_s\vert-1=-(1.28 \pm 0.28) 10^{-5}$.%
 % end of PACS codes
} %end of abstract
%
%\authorrunning
%\titlerunning
\maketitle
\section{Introduction}
\label{intro}
B physics plays an important role to test and improve our understanding of the Standard Model flavor-dynamics.

Theoretically, the large mass of the $b$ quark, compared to the QCD scale
parameter $\Lambda_{QCD}$, allows to treat inclusive rates in terms of an
operator product expansion (OPE), with a consequent separation of the
short-distance contributions from the long-distance ones.
Theoretical predictions of inclusive rates, therefore, are based on a non-perturbative calculation of matrix elements, widely studied in lattice QCD, and a perturbative calculation of Wilson coefficients.

Recently, the contribution of light quarks in beauty hadron decay widths (spectator effect) has been computed at $\mathcal{O}(\alpha_s)$ in QCD and $\mathcal{O}(\Lambda_{QCD}/m_b)$ in the OPE.
Based on these calculations is the theoretical prediction for both beauty hadron lifetimes and B-meson CP-violation parameters. 
Therefore, improved theoretical estimates have been
obtained, to be compared with recent accurate experimental measurements or
limits.
%%%%%%%%%%%%%%%%%%%%%%%%%%%%%%%%%%%%%%%%

\section{Beauty hadron lifetime ratios}
\label{sec:1}
%________________________________________________________________
\begin{figure*}
\begin{center}
\resizebox{0.5\textwidth}{!}{\includegraphics{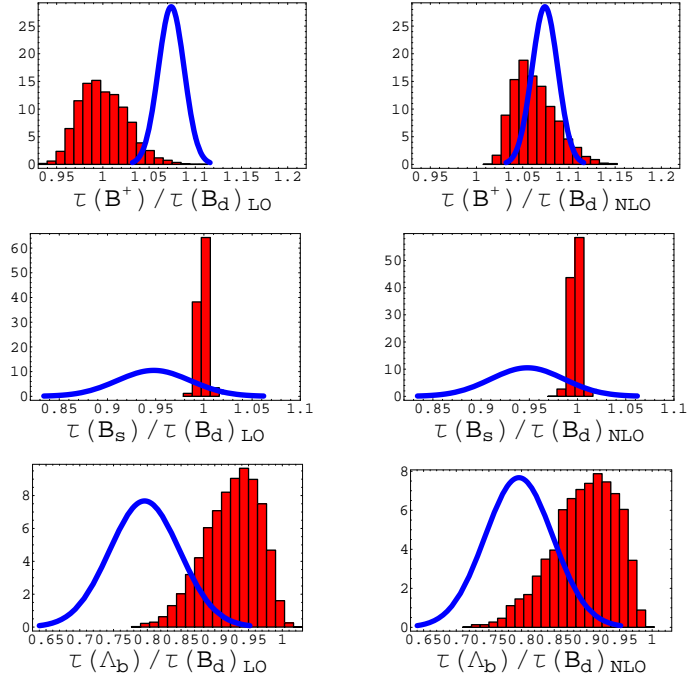}}
\end{center}
\caption{\it Theoretical (histogram) vs experimental (solid line) distributions of lifetime ratios. The theoretical predictions are shown at the LO (left) and NLO (right).}
\label{fig:plot}
\end{figure*}
The experimental values of the measured lifetime ratios of beauty hadrons are~\cite{BBpage}
\bea
\label{eq:rexp}
\frac{\tau(B^+)}{\tau(B_d)}=1.085 \pm 0.017 \, ,
\frac{\tau(B_s)}{\tau(B_d)}=0.951 \pm 0.038 \, ,\nn \\
\frac{\tau(\Lambda_b)}{\tau(B_d)}=0.786 \pm 0.034\, .\qquad\qquad\qquad
\eea

These quantities can be computed from first principles, since the great energy ($\sim m_b$) released in beauty hadron decays allows to expand the inclusive width $\Gamma(H_b)$ in powers of $1/m_b$, by applying the heavy quark expansion (HQE)~\cite{ope}.

Using the optical theorem, the inclusive decay width of a hadron $H_b$,
containing a $b$ quark, can be written as
\be
\Gamma(H_b) =
\frac{1}{M_{H_b}} {\mathrm Disc} \langle H_b \vert {\cal T} \vert H_b \rangle \,,
\label{eq:master}
\ee
where ``$\mathrm{Disc}$'' picks up the discontinuities across the physical cut in the transition operator $\cal T$, given by
\be 
{\cal T} = i \int d^4x \; T \left( \Heff^{\DB=1}(x) \Heff^{\DB=1}(0) \right)\, .\label{eq:T}
\ee
$\Heff^{\DB=1}$ is the effective weak Hamiltonian which describes 
$\DB=1$ transitions, whose Wilson coefficients are known at the next-to-leading order (NLO) in QCD~\cite{nlodb1a,nlodb1b,nlodb1c}. 

By applying the HQE, the decay width $\Gamma(H_b)$ in eq.~\ref{eq:master} can be expressed as  a sum of local $\DB=0$ operators of increasing dimension
\be
\Gamma (H_b) = \sum_k \frac{\vec{c}_k (\mu)}{m_b^k}\,\langle H_b | \vec{O}^{\DB = 0}_k (\mu) | H_b \rangle\nn\,.
\label{eq:sum}
\ee

The HQE brings to the separation of short distance effects, confined in the Wilson coefficients ($\vec{c}_k$) and evaluable in perturbation theory, from long distance physics, represented by the matrix elements of the local operators ($\vec{O}_k^{\DB=0}$), to be computed non-perturbatively.

Up to $\mathcal{O}(1/m_b^2)$, only the $b$ quark enters the short-distance weak decay, while the light spectator quarks, which distinguish different beauty hadrons, interact through soft gluons only.
The local operators appearing up to $\mathcal{O}(1/m_b^2)$ in the QCD HQE are the condensate ($\bar b b$) and the chromomagnetic operator ($\bar b \sigma_{\mu \nu} G^{\mu \nu} b$) which do not contain the light quark field. Their contribution can be evaluated from the heavy hadron spectroscopy and leads to the following estimates
\bea
\label{eq:mb2}
\frac{\tau(B^+)}{\tau(B_d)}=1.00 \, ,
\frac{\tau(B_s)}{\tau(B_d)}=1.00\, ,
\frac{\tau(\Lambda_b)}{\tau(B_d)}=0.98(1)\,, 
\eea
where the uncertainties on the first two ratios, being inferior to $1\%$, are not indicated.

Spectator contributions appear at $\mathcal{O}(1/m_b^3)$  in the HQE.
These effects, although suppressed by an additional power of $1/m_b$, are enhanced, with respect to leading contributions, by a phase-space factor of $16 \pi^2$, being $2 \to 2$ processes instead of $1 \to 3$ decays~\cite{Bigi:1992su,NS}.

In order to evaluate the spectator effects one has to calculate the matrix elements of dimension-six current-current and penguin operators, non-perturbatively, and their Wilson coefficients, perturbatively.

Last year both the non-perturbative and the perturbative calculations have been improved.

Concerning the perturbative part, the NLO QCD corrections to the coefficient functions of the current-current operators have been computed~\cite{Ciuchini:2001vx,BBB,NOI}.

Concerning the non-perturbative part, the usual parametrization of the matrix elements of the dimension-six current-current operators distinguishes two cases, depending on whether or not the light quark of the operator enters as a valence quark in the external hadronic state. Therefore, different B-parameters for the valence and non-valence contributions are introduced.
The reason for this parametrization is that so far the non-valence contributions have not been computed. Their non-perturbative lattice calculation would be possible, in principle, however it requires to deal with the difficult problem of power-divergence subtractions.
On the other hand, the valence contributions have been recently evaluated, for $B-$mesons, by combining the QCD and HQET lattice results to extrapolate to the physical $b$ quark mass~\cite{APE} and, for the $\Lambda_b$ baryon, in lattice-HQET~\cite{DiPierro:1999tb}.
These accurate results are in agreement with the values obtained in previous lattice studies~\cite{Becirevic:2001fy,DiPierro98,DiPierro:1998cj} and with the estimates based on QCD sum rules~\cite{Colangelo:1996ta,Baek:1998vk,Cheng:1999ia,Huang:1999xj}.

This year, the sub-leading spectator effects which appear at $\mathcal{O}(1/m_b^4)$ in the HQE, have been included in the analysis of lifetime ratios.
The relevant operator matrix elements have been estimated in the vacuum saturation approximation (VSA) for $B-$mesons and in the quark-diquark model for the $\Lambda_b$ baryon, while the corresponding Wilson coefficients have been calculated at the leading order (LO) in QCD~\cite{Gabbiani:2003pq}.

In this talk we update our theoretical predictions for the lifetime ratios~\cite{NOI}, which contain NLO QCD corrections to Wilson coefficients and lattice values for valence B-parameters, by including  the sub-leading spectator effects of ref.~\cite{Gabbiani:2003pq}. In this way we obtain
\bea
\label{eq:nlores}
\left. \frac{\tau(B^+)}{\tau(B_d)} \right|_{\rm NLO}\!\!\!\!\!\!\!\!\!\! =
1.06 \pm 0.02 \,,  
\left. \frac{\tau(B_s)}{\tau(B_d)} \right|_{\rm NLO}\!\!\!\!\!\!\!\!\!\! =
1.00 \pm 0.01 \,, \nn\\ 
\left. \frac{\tau(\Lambda_b)}{\tau(B_d)} \right|_{\rm NLO}\!\!\!\!\!\!\!\!\!\! =
0.88 \pm 0.05\,.\qquad\qquad\qquad
\eea
They turn out to be in good agreement with the experimental data of eq.~\ref{eq:rexp}.
It is worth noting that the agreement at $1.5 \sigma$ between the theoretical prediction for the ratio $\tau(\Lambda_b)/\tau(B_d)$ and its experimental value is achieved thanks to the inclusion of the NLO (see fig.~\ref{fig:plot}) and the $1/m_b$ corrections to spectator effects. 
They both decrease the central value of $\tau(\Lambda_b)/\tau(B_d)$ by $8\%$ and $2\%$ respectively.

Further improvement of the $\tau(\Lambda_b)/\tau(B_d)$ theoretical prediction would require the calculation of the current-current operator non-valence B-parameters and of the perturbative and non-perturbative contribution of the penguin operator, which appears at the NLO and whose matrix elements present the same problem of power-divergence subtraction.
These contributions are missing also in the theoretical predictions of $\tau(B^+)/\tau(B_d)$ and $\tau(B_s)/\tau(B_d)$, but in these cases they represent an effect of $SU(2)$ and $SU(3)$ breaking respectively, and are expected to be small.
%%%%%%%%%%%%%%%%%%%%%%%%%%%%%%%%%%%%%%%%

\section{Neutral $B_q$-meson width differences}
\label{sec:2}
The width difference between the ``light''  and ``heavy'' neutral $B_q$-meson ($q=d, s$) is defined in terms of the off-diagonal matrix element ($\Gamma^q_{21}$) of the absorptive part of the $B-\bar B$ mixing effective hamiltonian
\be
\Delta \Gamma_q \equiv \Gamma_L^q - \Gamma_H^q = -2 \Gamma_{21}^q \equiv - \frac{1}{M_{B_q}} {\mathrm Disc} \langle \overline{B}_q | \mathcal{T} | B_q \rangle \nn\,,
\label{eq:diff}
\ee
where the transition operator $\mathcal{T}$ is given in eq.~\ref{eq:T}.

As in the case of the inclusive decay widths discussed above, the great energy scale ($\sim m_b$) which characterizes the decay process allows the HQE of the amplitude in eq.~\ref{eq:diff} as a series of matrix elements of $\DB=2$ local operators, multiplied for their Wilson coefficients. 

The leading contribution comes at $\mathcal{O}(1/m_b^3)$ in the HQE and is given by the dimension-six $\DB=2$ operators.
Up to and including $\mathcal{O}(1/m_b^4)$ contribution, the HQE of $\Gamma_{21}^q$ reads

\bea
\Gamma^q_{21} &=&
-\dfrac{G_F^2 m_b^2}{24 \pi M_{B_q}}\left[
c^q_1(\mu_2) {\langle \overline B_q \vert {\cal O}^q_1(\mu_2) \vert B_q\rangle}
+ \right.\nn\\
& & \left.c^q_2(\mu_2) {\langle \overline B_q \vert {\cal O}^q_2(\mu_2) \vert B_q\rangle} 
+ \delta^q_{1/m}\right]\, ,
\label{eq:gamma12q}
\eea
where $c^q_i(\mu_2)$ are the Wilson coefficients, known at the NLO in QCD.
In the case of $\Delta \Gamma_s$ the NLO corrections have been computed in ref.~\cite{BBlargh}, whereas for $\Delta \Gamma_d$ the complete NLO corrections, including contributions from a non vanishing charm quark mass, have been calculated this year~\cite{Beneke:2003az,NOIwip}.
It is worth noting that the charm mass corrections are necessary to take fully into account the dependence of $\Delta \Gamma_d$ on the weak phase, contained in $c_i^d$, at the NLO accuracy.

${\langle \overline B_q \vert {\cal O}^q_i(\mu_2) \vert B_q\rangle}$ are the matrix elements of the two independent dimension-six operators, while $\hat{\delta}_{1 / m_b}$ represents the contribution of the dimension-seven operators~\cite{seven}.

Lattice results of the dimension-six operator matrix elements~\cite{Gimenez:2000jj}-{\cite{Lellouch:2000tw} have been confirmed and improved last year.
In order to reduce the systematics of the heavy quark extrapolation, the results obtained in QCD have been combined with the HQET ones~\cite{damir} (see fig.~\ref{fig:extr}).
The effect of the inclusion of the dynamical quarks has been examined, within the NRQCD approach, finding that these matrix elements are essentially insensitive to switching from $n_f=0$ to $n_f=2$~\cite{Yamada:2001xp,Aoki:2003xb}.
Lately, the same matrix elements have been calculated by using QCD sum rules with NLO accuracy~\cite{Korner:2003zk}, thus achieving a reduced uncertainty with respect to previous determinations in this theoretical framework~\cite{Hagiwara:2002hf,Narison:1994zt}.

%________________________________________________________________
\begin{figure*}
\begin{center}
\resizebox{0.5\textwidth}{!}{\includegraphics{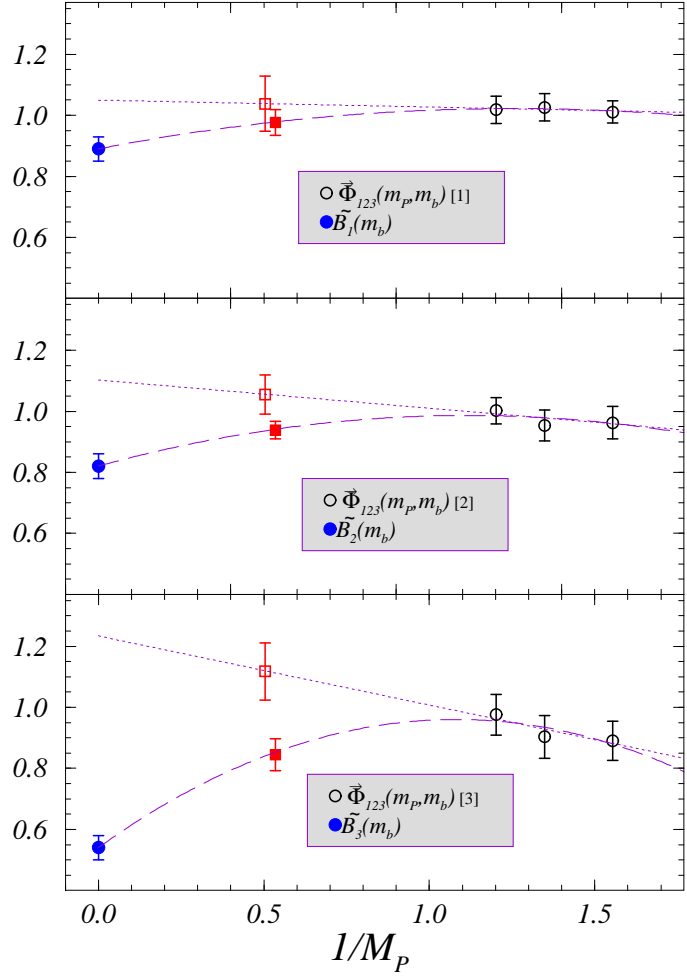}}
\end{center}
\caption{\it Extrapolation of $\Delta B=2$ B-parameters in the inverse heavy meson mass, by combining QCD ($\vec{\Phi}_{123}(m_P,m_b)$) and HQET ($\tilde{B}_i(m_b), i=1,2,3$) lattice results. The inclusion of the HQET point has the effect of decreasing the extrapolated value.}
\label{fig:extr}
\end{figure*}
%________________________________________________________________

On the other hand, the dimension-seven operator matrix elements have never been estimated out of the VSA.
However two of these four matrix elements can be related through Fierz identities to the complete set of operators studied in ref.~\cite{damir}.

The expression of $\Delta \Gamma_q /\Gamma_q$ used in the analysis, is obtained by neglecting $(\Gamma_{21}^q/M_{21}^q)^2=\mathcal{O}(m_b^4/m_t^4)$ terms ($M^q_{21}$ represents the off-diagonal matrix element of the dispersive part of the $B - \bar B$ mixing effective hamiltonian) and reads
\be
\label{dg}
\dfrac{\Delta \Gamma_q}{\Gamma_ q} = -\dfrac{\Delta M_q}{\Gamma_q}{\mathrm{Re}}\left(\dfrac{\Gamma^q_{21}}{M^q_{21}}\right)\,.
\ee

The updated theoretical predictions obtained in the analysis of ref.~\cite{NOIwip} are 
\be
\Delta \Gamma_{d}/\Gamma_{d} = (2.42 \pm 0.59)10^{-3} ,\,\,
\Delta \Gamma_{s}/\Gamma_{s} = (7.4 \pm 2.4)10^{-2} \,.
\ee
The corresponding theoretical distributions are shown in fig.~\ref{fig:plot1}, where the effect of the NLO corrections can be seen to be quite relevant.
%________________________________________________________________
\begin{figure*}
\begin{center}
\resizebox{0.6\textwidth}{!}{\includegraphics{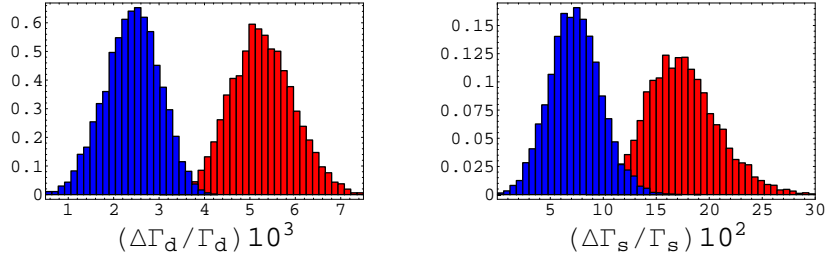}}
\end{center}
\caption{\it Theoretical distributions 
 for the width difference in $B_d$ and $B_s$ systems. 
The theoretical predictions are shown at the LO (light/red) and NLO (dark/blue).}
\label{fig:plot1}
\end{figure*}

One can see that $\Delta \Gamma_{s}$ is larger than $\Delta \Gamma_{d}$, the latter receiving contributions from channels which are doubly Cabibbo suppressed with respect to those contributing to $\Delta \Gamma_{s}$, and both agree with the experimental limits~\cite{BBpage}
\bea
\Delta \Gamma_{d}/\Gamma_{d} = 0.008 \pm 0.037 (stat.) \pm 0.019 (syst.)\,,\nn\\\Delta \Gamma_{s}/\Gamma_{s} = 0.07^{+0.09}_{-0.07}\,,\qquad\qquad
\eea
within the large experimental uncertainties.

In order to test the agreement between theoretical and experimental values with higher precision, it is important to wait for more accurate measurements from the $B-$factories (Babar and Belle)  and from the RunII at Tevatron and the LHC.
%%%%%%%%%%%%%%%%%%%%%%%%%%%%%%%%%%%%%%%%
\section{CP Violation parameters: $\vert(q/p)_d\vert$ and $\vert(q/p)_s\vert$}
\label{sec:3}
%%%%%%%%%%%%%%%%%%%%%%%%%%%%%%%%%%%%%%%%
%________________________________________________________________
\begin{figure*}
\begin{center}
\resizebox{0.6\textwidth}{!}{\includegraphics{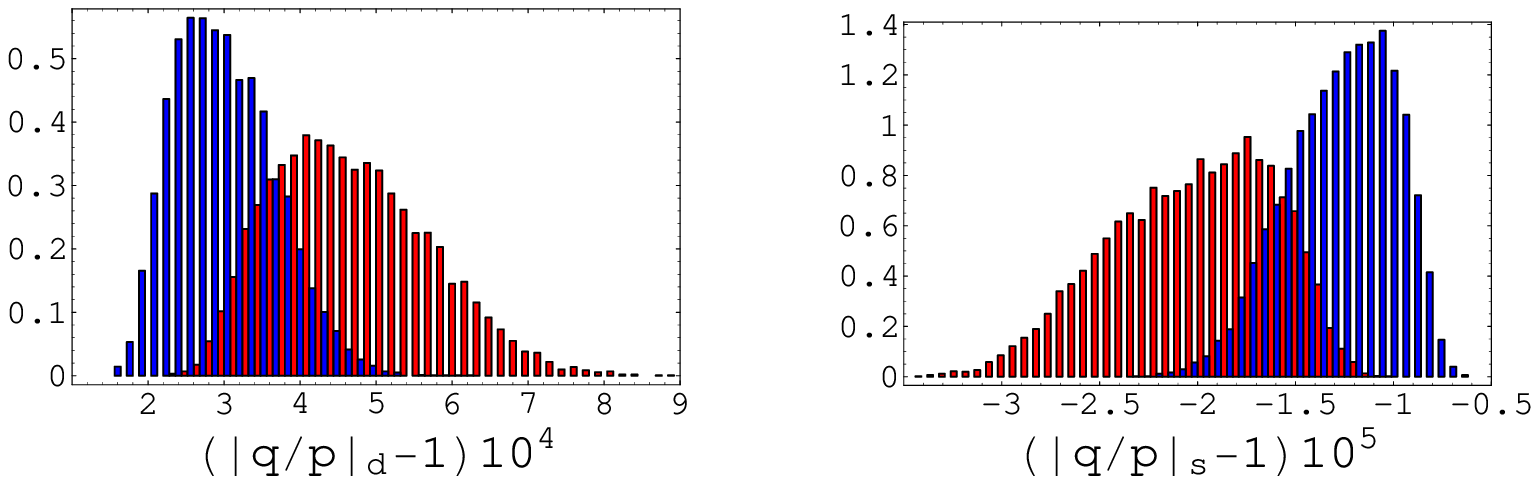}}
\end{center}
\caption{\it Theoretical distributions 
 for $\vert(q/p)_q\vert-1$ in $B_d$ and $B_s$ systems. 
The theoretical predictions are shown at the LO (light/red) and NLO (dark/blue).}
\label{fig:plot3}
\end{figure*}
%________________________________________________________________
The experimental observable $\vert(q/p)_q\vert$, whose deviation from unity describes CP-violation due to mixing, is related to $M_{21}^q$ and $\Gamma_{21}^q$, through
\be
\left({q/p}\right)_q=\sqrt{
\dfrac{2 M^q_{21}-i\Gamma^q_{21}}
{2 M^{q\,*}_{21}-i\Gamma^{q\,*}_{21}}}\,,
\ee
which, neglecting $(\Gamma_{21}^q/M_{21}^q)^2=\mathcal{O}(m_b^4/m_t^4)$ terms, becomes
\be
\label{eq:qsp}
\left\vert\left(\dfrac{q}{p}\right)_q\right\vert=1+
\dfrac{1}{2}\,{\mathrm{Im}}\left(\dfrac{\Gamma^q_{21}}{M^q_{21}}\right)\,.
\ee

By comparing eqs.~\ref{dg} and~\ref{eq:qsp}, one sees that the theoretical prediction of $\vert(q/p)_q\vert$ is based on the same perturbative and non perturbative calculation discussed in section~\ref{sec:2}.
In other words $\vert(q/p)_q\vert -1$ differs from $\Delta \Gamma_q /\Gamma_q$, a part from multiplicative factors, for the presence of ``${\mathrm{Im}}$'' instead of ``${\mathrm{Re}}$'', which selects a different contribution from $V_{CKM}$.

An important consequence of different CKM contributions is that $(\vert(q/p)_q\vert-1)/\Delta \Gamma_q={\mathcal O}(m_c^2/m_b^2)$. In the limit $m_c \rightarrow 0$, indeed, there are two quarks ($u$ and $c$) with the same charge and degenerate in mass, so that one can eliminate the CP-violating phase from $V_{CKM}$, through a quark field redefinition.

Moreover we have $(\vert(q/p)_s\vert-1)/(\vert(q/p)_d\vert-1)={\mathcal O}(\lambda^2)$ (where $\lambda$ is the sine of the Cabibbo angle) and, quantitatively, we find the updated theoretical predictions~\cite{NOIwip}
\bea
\label{eq:nlores3}
\vert(q/p)_d\vert-1=(2.96 \pm 0.67) 10^{-4}\,,\nn\\
\vert(q/p)_s\vert-1=-(1.28 \pm 0.28) 10^{-5}\,.
\eea
The corresponding theoretical distributions are shown in fig.~\ref{fig:plot3}.
Also for these quantities, the effect of NLO corrections turns out to be rather important.

A preliminary measurement for $\vert(q/p)_d\vert -1$  is now available from the BABAR collaboration~\cite{Aubert:2003se}
\be  
\vert (q/p)_d\vert -1=0.029\pm 0.013(\rm{stat.})\pm 0.011(\rm{syst.})
\ee
Improved measurements are certainly needed to make this comparison more significant.

\vspace*{1.0cm}

It is a pleasure to thank D.~Becirevic, M.~Ciuchini, E.~Franco, V.~Lubicz and F.~Mescia for sharing their insights in topics covered by this talk. I also thank the EPS-2003 organizers for the very stimulating conference realized in Aachen.

%%%%%%%%%%%%%%%%%%%%%%%%%%%%%%%%%%%%%%%% 

% BibTeX users please use
% \bibliographystyle{}
% \bibliography{}

\begin{thebibliography}{}

\bibitem{BBpage}
M.~Battaglia {\it et al},
%``The CKM matrix and the unitarity triangle,''
hep-ph/0304132.\\
%%CITATION = HEP-PH 0304132;%%
See also The Heavy Flavor Averaging Group (HFAG),
http://www.slac.stanford.edu/ xorg/hfag/


\bibitem{ope}
J.~Chay {\it et al.},
%``Lepton Energy Distributions In Heavy Meson Decays From QCD,''
Phys.\ Lett.\ B {\textbf 247} (1990) 399.
%%CITATION = PHLTA,B247,399;%%

\bibitem{nlodb1a}
%\cite{Buras:1993tc}
A.J.~Buras {\it et al.},
%``Two loop anomalous dimension matrix for Delta S = 1 weak nonleptonic decays. 1. O(alpha-s**2),''
Nucl.\ Phys.\ B {\textbf 400} (1993) 37
[hep-ph/9211304];
%%CITATION = HEP-PH 9211304;%%
%\cite{Buras:1993zv}

\bibitem{nlodb1b}
A.J.~Buras {\it et al.},
%``Two loop anomalous dimension matrix for Delta S = 1 weak nonleptonic decays. 2. O(alpha-alpha-s),''
Nucl.\ Phys.\ B {\textbf 400} (1993) 75
[hep-ph/9211321].
%%CITATION = HEP-PH 9211321;%%
%\cite{Ciuchini:1994vr}

\bibitem{nlodb1c}
M.~Ciuchini  {\it et al.},
%``The Delta S = 1 effective Hamiltonian including next-to-leading order QCD and QED corrections,''
Nucl.\ Phys.\ B {\textbf 415} (1994) 403
[hep-ph/9304257].
%%CITATION = HEP-PH 9304257;%%

%\cite{Bigi:1992su}
\bibitem{Bigi:1992su}
I.~I.~Bigi  {\it et al.},
%``Nonperturbative corrections to inclusive beauty and charm decays: QCD versus phenomenological models,''
Phys.\ Lett.\ B {\bf 293} (1992) 430
[Erratum-ibid.\ B {\bf 297} (1993) 477]
[hep-ph/9207214].
%%CITATION = HEP-PH 9207214;%%

\bibitem{NS}
M.~Neubert and C.T.~Sachrajda,
%``Spectator effects in inclusive decays of beauty hadrons,''
Nucl.\ Phys.\ B {\textbf 483} (1997) 339
[hep-ph/9603202].

%\cite{Ciuchini:2001vx}
\bibitem{Ciuchini:2001vx}
M.~Ciuchini {\it et al.},
%``Next-to-leading order QCD corrections to spectator effects in lifetimes  of beauty hadrons,''
Nucl.\ Phys.\ B {\bf 625} (2002) 211
[hep-ph/0110375].
%%CITATION = HEP-PH 0110375;%%

%\cite{Beneke:2002rj}
\bibitem{BBB}
M.~Beneke {\it et al.},
%``The B+ - B/d0 lifetime difference beyond leading logarithms,''
Nucl.\ Phys.\ B {\bf 639} (2002) 389
[hep-ph/0202106].
%%CITATION = HEP-PH 0202106;%%

%\cite{Franco:2002fc}
\bibitem{NOI}
E.~Franco {\it et al.},
%``Lifetime ratios of beauty hadrons at the next-to-leading order in QCD,''
Nucl.\ Phys.\ B {\textbf 633} (2002) 212
[hep-ph/0203089].
%%CITATION = HEP-PH 0203089;%%

\bibitem{APE} D.~Becirevic, private communication; updated results with respect
to hep-ph/0110124

%\cite{DiPierro:1999tb}
\bibitem{DiPierro:1999tb}
M.~Di Pierro {\it et al.} [UKQCD collaboration],
%``An exploratory lattice study of spectator effects in inclusive decays  of the Lambda/b baryon,''
Phys.\ Lett.\ B {\textbf 468} (1999) 143
[hep-lat/9906031].
%%CITATION = HEP-LAT 9906031;%%

\bibitem{Becirevic:2001fy}
D.~Becirevic,
%``Theoretical progress in describing the B meson lifetimes,''
hep-ph/0110124.
%%CITATION = HEP-PH 0110124;%%

%\cite{DiPierro:1998ty}
\bibitem{DiPierro98}
M.~Di Pierro and C.T.~Sachrajda  [UKQCD Collaboration],
%``A lattice study of spectator effects in inclusive decays of B mesons,''
Nucl.\ Phys.\ B {\textbf 534} (1998) 373
[hep-lat/9805028].
%%CITATION = HEP-LAT 9805028;%%M.~Di Pierro, C.~T.~Sachrajda, mesoni B


\bibitem{DiPierro:1998cj}
M.~Di Pierro and C.~T.~Sachrajda  [UKQCD collaboration],
%``Spectator effects in inclusive decays of beauty hadrons,''
Nucl.\ Phys.\ Proc.\ Suppl.\  {\textbf 73} (1999) 384
[hep-lat/9809083].
%%CITATION = HEP-LAT 9809083;%%

%\cite{Colangelo:1996ta}
\bibitem{Colangelo:1996ta}
P.~Colangelo and F.~De Fazio,
%``Role of four-quark operators in the inclusive Lambda/b decays,''
Phys.\ Lett.\ B {\textbf 387} (1996) 371
[hep-ph/9604425].
%%CITATION = HEP-PH 9604425;%%

%\cite{Baek:1998vk}
\bibitem{Baek:1998vk}
M.~S.~Baek {\it et al.},
% %``Four-quark operators relevant to B meson lifetimes from {QCD} sum rules,''
Phys.\ Rev.\ D {\textbf 57}, 4091 (1998) [hep-ph/9709386].
%%CITATION = HEP-PH 9709386;%%

%\cite{Cheng:1999ia}
\bibitem{Cheng:1999ia}
H.~Y.~Cheng and K.~C.~Yang,
%``Nonspectator effects and B meson lifetimes from a field-theoretic  calculation,''
Phys.\ Rev.\ D {\textbf 59} (1999) 014011
[hep-ph/9805222].
%%CITATION = HEP-PH 9805222;%%

\bibitem{Huang:1999xj}
C.~S.~Huang {\it et al.},
%``Reanalysis of the four-quark operators relevant to Lambda/b lifetime from {QCD} sum rule,''
Phys.\ Rev.\ D {\textbf 61}, 054004 (2000)
[hep-ph/9906300].
%%CITATION = HEP-PH 9906300;%%

%\cite{Gabbiani:2003pq}
\bibitem{Gabbiani:2003pq}
F.~Gabbiani {\it et al.},
%``Lambda/b lifetime puzzle in heavy-quark expansion,''
hep-ph/0303235.
%%CITATION = HEP-PH 0303235;%%

%\cite{Beneke:1998sy}
\bibitem{BBlargh}
M.~Beneke {\it et al.},
%``Next-to-leading order {QCD} corrections to the lifetime difference of  B/s mesons,''
Phys.\ Lett.\ B {\textbf 459} (1999) 631
[hep-ph/9808385].
%%CITATION = HEP-PH 9808385;%%

\bibitem{Beneke:2003az}
M.~Beneke {\it et al.},
%``CP asymmetry in flavour-specific B decays beyond leading logarithms,''
hep-ph/0307344.
%%CITATION = HEP-PH 0307344;%%

\bibitem{NOIwip}
M.~Ciuchini {\it et al.},
hep-ph/0308029.

\bibitem{seven}
M.~Beneke {\it et al.},
%``Width Difference in the $B_s-\bar{B_s}$ System,''
Phys.\ Rev.\ D {\textbf 54} (1996) 4419
[hep-ph/9605259].
%%CITATION = HEP-PH 9605259;%%

%\cite{Gimenez:2000jj}
\bibitem{Gimenez:2000jj}
V.~Gimenez and J.~Reyes,
%``Quenched and first unquenched lattice HQET determination of the  B/s meson width difference,''
Nucl.\ Phys.\ Proc.\ Suppl.\  {\textbf 94} (2001) 350
[hep-lat/0010048].
%%CITATION = HEP-LAT 0010048;%%

%\cite{Hashimoto:2000eh}
\bibitem{Hashimoto:2000eh}
S.~Hashimoto {\it et al.},
%``Renormalization of the Delta(B) = 2 four-quark operators in lattice  NRQCD,''
Phys.\ Rev.\ D {\textbf 62} (2000) 114502
[hep-lat/0004022].
%%CITATION = HEP-LAT 0004022;%%

%\cite{Aoki:2002bh}
\bibitem{Aoki:2002bh}
S.~Aoki {\it et al.} [JLQCD Collaboration],
%``B0 - anti-B0 mixing in quenched lattice QCD,''
Phys.\ Rev.\ D {\textbf 67} (2003) 014506
[hep-lat/0208038].
%%CITATION = HEP-LAT 0208038;%%

%\cite{Becirevic:2000sj}
\bibitem{Becirevic:2000sj}
D.~Becirevic {\it et al.},
%``A theoretical prediction of the B/s meson lifetime difference,''
Eur.\ Phys.\ J.\ C {\textbf 18} (2000) 157
[hep-ph/0006135].
%%CITATION = HEP-PH 0006135;%%

\bibitem{Lellouch:2000tw}
L.~Lellouch and C.~J.~Lin  [UKQCD Collaboration],
%``Standard model matrix elements for neutral B meson mixing and  associated decay constants,''
Phys.\ Rev.\ D {\textbf 64} (2001) 094501
[hep-ph/0011086].
%%CITATION = HEP-PH 0011086;%%

%\cite{Becirevic:2001xt}
\bibitem{damir}
D.~Becirevic {\it et al.},
%``B-parameters of the complete set of matrix elements of Delta(B) = 2  operators from the lattice,''
JHEP {\textbf 0204} (2002) 025
[hep-lat/0110091].
%%CITATION = HEP-LAT 0110091;%%


%\cite{Yamada:2001xp}
\bibitem{Yamada:2001xp}
N.~Yamada {\it et al.} [JLQCD Collaboration],
%``B meson B-parameters and the decay constant in two-flavor dynamical  QCD,''
Nucl.\ Phys.\ Proc.\ Suppl.\  {\textbf 106} (2002) 397
[hep-lat/0110087].
%%CITATION = HEP-LAT 0110087;%%

%\cite{Aoki:2003xb}
\bibitem{Aoki:2003xb}
S.~Aoki {\it et al.} [JLQCD Collaboration],
%``B0 anti-B0 mixing in unquenched lattice QCD,''
hep-ph/0307039.
%%CITATION = HEP-PH 0307039;%%

%\cite{Korner:2003zk}
\bibitem{Korner:2003zk}
J.~G.~Korner {\it et al.},
%``B0 anti-b0 mixing beyond factorization,''
hep-ph/0306032.
%%CITATION = HEP-PH 0306032;%%

%\cite{Hagiwara:2002hf}
\bibitem{Hagiwara:2002hf}
K.~Hagiwara {\it et al.},
%``B0/d,s - anti-B0/d,s mass-differences from QCD spectral sum rules,''
Phys.\ Lett.\ B {\bf 540}, 233 (2002)
[hep-ph/0205092].
%%CITATION = HEP-PH 0205092;%%

%\cite{Narison:1994zt}
\bibitem{Narison:1994zt}
S.~Narison and A.~A.~Pivovarov,
%``QSSR estimate of the B(B) parameter at next-to-leading order,''
Phys.\ Lett.\ B {\bf 327}, 341 (1994)
[hep-ph/9403225].
%%CITATION = HEP-PH 9403225;%%

%\cite{Aubert:2003se}
\bibitem{Aubert:2003se}
B.~Aubert {\it et al.} [BABAR Collaboration],
%``Limits on the lifetime difference of neutral B mesons and on CP, T, and  CPT violation in B0 anti-B0 mixing,''
hep-ex/0303043.
%%CITATION = HEP-EX 0303043;%%


\end{thebibliography}

\end{document}